\documentclass[aps,prb,reprint,showpacs,preprintnumbers,superscriptaddress,amsmath,amssymb]{revtex4-1}

\usepackage{ae,aecompl}
\usepackage{graphicx}
\usepackage{epstopdf}
\usepackage[version=3]{mhchem}
\usepackage{float}
\usepackage[dvips]{hyperref}
\usepackage{textcomp}
\usepackage[OT1,T1]{fontenc}

\usepackage{color}

\begin{document}

\title{Orthorhombic fulleride \ce{(CH3NH2)K3C60} close to Mott-Hubbard instability: \textit{Ab initio} study}

\author{Anton Poto\v{c}nik}
\email{anton.potocnik@ijs.si}
\affiliation{Condensed Matter Physics Department, Jo\v{z}ef Stefan Institute, Jamova cesta 39, SI-1000 Ljubljana, Slovenia.}

\author{Nicola Manini}
\email{nicola.manini@mi.infm.it}
\affiliation{Dipartimento di Fisica, Universit\`{a} degli Studi di Milano, via Celoria 16, 20133 Milano, Italy.}
\affiliation{Scuola Internazionale Superiore di Studi Avanzati, Via Bonomea 265, 34136 Trieste, Italy.}
\affiliation{CNR-IOM Democritos, Via Bonomea 265, 34136 Trieste, Italy.}

\author{Matej Komelj}
\affiliation{Condensed Matter Physics Department, Jo\v{z}ef Stefan Institute, Jamova cesta 39, SI-1000 Ljubljana, Slovenia.}

\author{Erio Tosatti}
\email{erio.tosatti@sissa.it}
\affiliation{Scuola Internazionale Superiore di Studi Avanzati, Via Bonomea 265, 34136 Trieste, Italy.}
\affiliation{CNR-IOM Democritos, Via Bonomea 265, 34136 Trieste, Italy.}
\affiliation{International Centre for Theoretical Physics (ICTP), Strada Costiera 11, 34151 Trieste, Italy.}

\author{Denis Ar\v{c}on}
\affiliation{Condensed Matter Physics Department, Jo\v{z}ef Stefan Institute, Jamova cesta 39, SI-1000 Ljubljana, Slovenia.}
\affiliation{Faculty of mathematics and physics, University of Ljubljana, Jadranska 19, SI-1000 Ljubljana, Slovenia.}

\date{\today}

\begin{abstract}
We study the electronic structure and magnetic interactions in methylamine-intercalated orthorhombic alkali-doped fullerene \ce{(CH3NH2)K3C60} within the density functional theory. As in the simpler ammonia intercalated compound \ce{(NH3)K3C60}, the orthorhombic crystal-field anisotropy $\mathit{\Delta}$ lifts the $t_\mathrm{1u}$ triple degeneracy at the $\Gamma$  point and drives the system deep into the Mott-insulating phase. However, the computed $\mathit{\Delta}$ and conduction electron bandwidth \textit{W} cannot alone account for the abnormally low experimental N\'eel temperature, $T_{\rm{N}} = 11$~K of the methylamine compound, compared to the much higher value $T_{\rm{N}} = 40$~K of the ammonia one. Significant interactions between \ce{CH3NH2} and \ce{C60^{3-}} are responsible for the stabilization of particular pseudo-Jahn-Teller fullerene-cage distortions and the ensuing low-spin $S = 1/2$ state. These interactions also seem to affect the magnetic properties, as interfullerene exchange interactions depend on the relative orientation of pseudo-Jahn-Teller distortions of neighboring \ce{C60^{3-}} molecules. For the ferro-orientational order of \ce{CH3NH2}-K$^+$ groups we find an apparent reduced dimensionality in magnetic exchange interactions, which may explain the suppressed N\'eel temperature. The disorder in exchange interactions caused by orientational disorder of \ce{CH3NH2}-K$^+$ groups could further contribute to this suppression. 
\end{abstract}

\maketitle

\section{Introduction}
The alkali-doped cubic fullerides ($A_{3-x}A'_{x}$\ce{C60}, $A$, $A'$ = Na, K, Rb, Cs) are prominent members of a rapidly growing family of $\pi$-electron organic superconductors \cite{Mitsuhashi} with superconducting transition temperature, $T_c$, as high as 38~K.\cite{Ganin2008,Pennington} In the early days of fullerene research the observed monotonic increase of $T_c$ as a function of unit cell volume was attributed to the increased density of states at the Fermi level, within a standard Bardeen-Cooper-Schrieffer (BCS) theory.\cite{Gunnarsson1997} Due to the three-fold degeneracy of the $t_\mathrm{1u}$-derived bands at the  $\Gamma$ point, preserved in cubic structures, and to the well-established vibron coupling, the Jahn-Teller (JT) effect is believed to participate actively in the superconducting pairing mechanism.\cite{Fabrizio1997} However, since the conduction electron's bandwidth ($W$) and the on-site Coulomb repulsion energy ($U$) are comparable it was also argued that fullerides must be very close to a Mott-Hubbard transition.\cite{Lof} Phase transitions from a superconducting to a Mott insulating state were indeed demonstrated long ago upon intercalation-induced lattice expansion in compounds such as \ce{(NH3)K3C60}\cite{Iwasa} where, similarly to \ce{Li3(NH3)6C60},\cite{Durand} expansion increases the repulsion/bandwidth ratio $U/W$ raising the importance of correlations. These notions recently received strong support with the discovery of \ce{Cs3C60}, a cubic fulleride with the largest unit cell and low-spin ($S = 1/2$) antiferromagnetic Mott-insulating ground state at ambient pressure conditions\cite{Ganin2008,Ganin2010,Takabayashi2009,Jeglic,Ihara2010,Ihara2011} and where high-temperature superconductivity is revived under hydrostatic pressure with no structural symmetry change. Strikingly, under pressure here $T_c$ displays a dome-like unit-cell volume dependence typical of unconventional superconductors,\cite{Chu} frankly inexplicable within BCS, Migdal-Eliashberg and related weakly-interacting models. Conversely, precisely such a nonmonotonic behavior of $T_c$ had been predicted by dynamical mean field theory (DMFT) for the unconventional superconducting phase bordering a Mott-insulating state at large unit cell volumes in a simple three-band Hubbard model incorporating, besides a JT electron-vibron coupling, strong on-site electron correlations, caused both by the Coulomb repulsion $U$ and by Hund's rule exchange, $J_\mathrm{H}$.\cite{Capone} That success not only highlights the unconventional interplay of strong correlations and phonons in these systems: it also brings expanded alkali-doped fullerenes at the forefront of the strongly-correlated electron systems and superconductors, where good understanding of one model system may help shed more light on the entire highly controversial field.

In search for higher transition temperatures, earlier approaches toward expanded unit cells had been by means of co-intercalation of inert ammonia (\ce{NH3}), e.g. \ce{NH3K_{3-x}Rb_{x}C60} ($x$ = 0, 1, 2, 3)\cite{Rosseinsky,Takenobu} and in \ce{Li3(NH3)6C60},\cite{Durand} or of methylamine (MA) molecules, in \ce{(CH3NH2)K3C60} (\ce{MAK3C60}).\cite{Ganin2006}  These molecules operate as spacers between \ce{C60} molecules and are believed to have negligible direct influence on the electronically active $t_\mathrm{1u}$ molecular orbitals of \ce{C60}$^{3-}$ anions. However, their presence breaks the original cubic symmetry resulting in orthorhombic crystal structures and effectively lifting the three-fold degeneracy of the $t_\mathrm{1u}$-derived bands at the $\Gamma$ point. This splitting reduces the critical value for $U/W_c$,\cite{Gunnarsson1996} immediately pushing non-cubic fullerides over the metal-insulator transition (MIT) boundary.\cite{Iwasa,Manini2002} We stress that these anisotropic compounds would still be metals, presumably also BCS superconductors, if strong on-site electron correlations were not present.\cite{Manini2002} They are thus important model systems for investigations of correlation effects in metals where the splitting of the originally degenerate narrow bands can be tuned through the anisotropy field. 

At low temperatures \ce{NH3K_{3-x}Rb_xC60} compounds order to an antiferromagnetic insulating state with N\'eel temperature, $T_\mathrm{N}$, ranging from 40~K to the maximum of 80~K for x = 0 and x = 2, respectively.\cite{Rosseinsky,Prassides} \ce{MAK3C60}, on the other hand, orders to an antiferromagnetic state at remarkably lower $T_\mathrm{N}$ = 11~K.\cite{Takabayashi2006} Assuming that the electronic properties depend mainly on the unit-cell volume one would anticipate $T_\mathrm{N}$ to be  $\sim$80~K, i.e. similar to that of \ce{NH3KRb2C60}. Such profoundly different magnetic response to different co-intercalands is unexpected and implies that in addition to direct interfullerene electronic overlap also other degrees of freedom play a role in non-cubic fullerides. We first note that the $c/a$ lattice parameter ratio is smaller in \ce{MAK3C60} than in \ce{NH3K_{3-x}Rb_xC60}, where it is closer to unity.\cite{Rosseinsky,Ganin2006} Therefore, the difference between the two orthorhombic compounds may arise from the different orthorhombic anisotropy field, $\mathit\Delta$. In other words, properties of \ce{MAK3C60} and \ce{NH3K_{3-x}Rb_xC60} compounds are fine-tuned by the closeness to the Mott-Hubbard boundary on the ($U,\mathit\Delta$) Manini-Santoro-Dal Corso-Tosatti (MSDT) phase diagram for fullerides.\cite{Manini2002} While it is well established that \ce{NH3K3C60} lays very close to this boundary, the precise position of \ce{MAK3C60} is yet to be determined.

In cubic fullerides the high crystal symmetry prevents ordering of the $t_\mathrm{1u}$ molecular orbitals, i.e. it is believed that cubic $A_3\mathrm{C}_{60}$ are in an orbitally liquid state,\cite{Gunnarsson1997} at least above the antiferromagnetic N\'eel temperature. In the non-cubic fullerides, however, one expects that crystal field will select a particular \ce{C60^{3-}} cage distortion and thus would also affect the inter-fulleride hopping integrals. Since MA molecule is rather large compared to available octahedral space in the fulleride structure a strong interaction - presumably even the formation of a weak hydrogen bond - between MA-K$^+$ group and the nearest \ce{C60} was proposed.\cite{Ganin2006} \ce{MAK3C60} thus offers a unique opportunity to investigate the role of symmetry breaking on the JT effect in strongly-correlated electron systems. Stimulated by these open issues we decided to carry out a systematic \textit{ab initio} study of \ce{MAK3C60} by the density functional theory (DFT) in the local density approximation (LDA), where even if strong correlations are treated only at the mean-field level, the detailed chemical bonding and crystal-field strengths can be assessed. We show that the crystal-field anisotropy of \ce{MAK3C60} is larger compared to \ce{NH3K3C60}, thus placing it deeper in the insulating region of the phase diagram. The whole electron-hopping structure is found to be significantly affected by the MA insertion. We find evidences for the presence of strong \ce{C60^{3-}} pseudo-JT effect that may lead to a reduced dimensionality in magnetic exchange interactions and explain suppressed $T_\mathrm{N}$.

\section{Methods}
We execute DFT calculations using the Quantum Espresso software package (pwscf program).\cite{Giannozzi} Ultrasoft pseudo potentials appropriate for the Perdew-Zunger exchange-correlation (LDA) are used. The pseudo Bloch functions are expanded over plane waves with an energy cutoff of 50~Ry on a $5^3$ Monkhorst-Pack \textit{k}-space mesh, such that the total energy is within 250~meV of its converged value, while the conduction bands are within 0.1~meV of their converged value. The Fermi surface is smeared with a ``temperature" parameter of 2~meV. The density of states (DOS) is evaluated with the tetrahedron method, sampling \textit{k}-space with a uniform $8^3$ mesh. For the relaxation of the atomic positions, we adopt a damp (quick-min Verlet) procedure on a $2^3$ \textit{k}-space mesh and plane-waves' energy cutoff of 30~Ry. The maximally-localized Wannier orbitals are obtained using the Wannier90 package.\cite{Mostofi} The three Wannier orbitals are computed in the $t_\mathrm{1u}$ energy window, using $5^3$ \textit{k}-points on Monkhorst-Pack grid with Bloch phases as initial projections. All three Wannier orbitals are positioned on the central \ce{C60} molecule and have a similar spread of  $\sqrt{\mathit{\Omega}} \sim 0.4$~nm. The projected DOS (PDOS) on the three Wannier orbitals were evaluated using the Gaussian smearing method with a ``temperature" parameter of 8~meV. The use of two different methods for evaluation of DOS and PDOS gives rise to a slight mismatch between the two. The mismatch does not affect the conclusions of this work.

\section{Results and Discussion}

\subsection{Phase diagram}
The \ce{MAK3C60} compound grows in the face-centered orthorhombic (space group \textit{Fmmm}) crystal structure with the room-temperature unit-cell parameters: $a = 15.2027$~\AA, $b = 15.1800$~\AA~and $c = 13.5032$~\AA~($V = 779.057$~\AA$^3$).\cite{Ganin2006} The MA-K$^+$ groups situated only in the large octahedral sites were found to be dynamically disordered between eight equivalent orientations at high temperatures.\cite{Arcon} Below the structural phase transition at $T_s = 220$~K MA-K$^+$ groups become static and probably ordered in an (anti)ferro-orientational order similar to \ce{NH3K3C60}.\cite{Margadonna} Since the details of the low temperature structure are not known yet we base all our band-structure calculations on the experimental room-temperature structure. We select a single MA-K$^+$ orientation thus effectively imposing a ferro-orientational order of the MA-K$^+$ groups.

\begin{figure}
\includegraphics[trim = 3mm 18mm 25mm 25mm, clip, width=8.6cm]{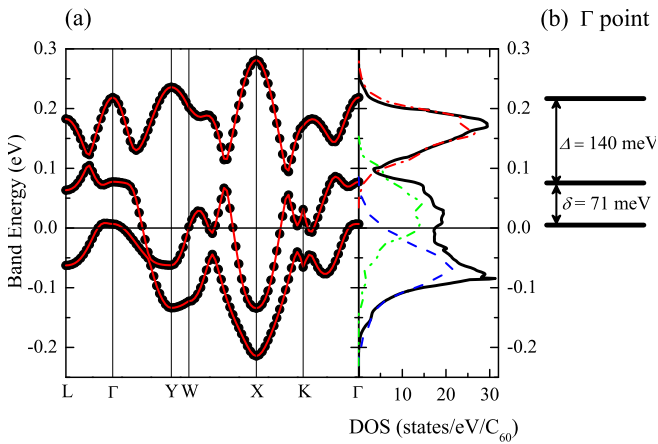}
\caption{(a) Band-structure and DOS for the room temperature \ce{MAK3C60} crystal structure. Red solid line is the interpolated band-structure using the maximally localized Wannier orbitals - see text for details. Black solid line: the total DOS; the red dot-dashed line, green dash-dot-dotted line, and blue dashed line represent the projected DOS on the first, second and the third Wannier orbital, respectively. (b) The splitting of $t_\mathrm{1u}$ bands at the $\Gamma$ point, defining the orthorhombic crystal-field anisotropy.}
\label{fig-1}
\end{figure}

The three metallic $t_\mathrm{1u}$ bands are well isolated from other molecular orbital-derived bands as is the case with every fullerene compound.\cite{Gunnarsson2004} The band gap between ``\ce{C60} HOMO" ($h_\mathrm{u}$-derived bands) and ``\ce{C60} LUMO" ($t_\mathrm{1u}$-derived bands) is 0.96~eV, while the separation between ``\ce{C60} LUMO" and ``\ce{C60} LUMO+1" ($t_\mathrm{1g}$-derived bands) is 0.59~eV. For comparison we refer here to the corresponding gaps of 1.16~eV and 0.38~eV computed for cubic \ce{K3C60},\cite{Haddon} and to the experimental values, roughly 1.8~eV and 1~eV respectively.\cite{Macovez} The underestimation of the band gaps, standard for LDA calculations, is fortunately of little consequence in our case. The neat separation of the different band groups allows us to focus entirely on $t_\mathrm{1u}$-derived bands close to the Fermi energy (Fig.~\ref{fig-1}). Not surprisingly, bare DFT-LDA nonmagnetic calculations yield for \ce{MAK3C60} a metallic ground state with a half-filled $t_\mathrm{1u}$ band. However, several experiments proved that \ce{MAK3C60} is an insulator,\cite{Takabayashi2006,Arcon,Arcon2008,Ganin2007} clearly suggesting that the mean-field DFT metallic state is driven to Mott-Hubbard insulating state by electron correlations. Since the phase diagram for non-cubic fullerides has already been calculated within DMFT with included electron correlations, it is sufficient at this stage to continue the characterization of \ce{MAK3C60} in the non-correlated limit and then treat electron correlations, e.g. within the established MSDT phase diagram.\cite{Manini2002}

Orthorhombic crystal symmetry is reflected in the band-structure [Fig.~\ref{fig-1}(a)] by (i) the removal of the threefold degeneracy at the $\Gamma$ point [Fig.~\ref{fig-1}(b)]; and (ii) the inequivalence of Y and X points. At the $\Gamma$ point we compute an energy splitting between the first and the second and between the second and the third $t_\mathrm{1u}$ energy levels of $\mathit{\delta} = 71$~meV and $\mathit{\Delta} = 140$~meV, respectively. For comparison, the corresponding energy splittings for \ce{NH3K3C60} are $\mathit{\delta} \approx 30$~meV and $\mathit{\Delta} \approx 150$~eV.\cite{Manini2002} In general, $\mathit\Delta$ reflects the major orthorhombic crystal-field anisotropy and $\mathit{\delta}$ tracks the smaller anisotropy in the $ab$ plane. Surprisingly, band-structure anisotropies of the two systems are very much comparable despite some obvious crystallographic differences, like for instance the ratio $c/a$, which is 0.89 and 0.91 for \ce{MAK3C60} and \ce{NH3K3C60}, respectively. In order to position \ce{MAK3C60} on the MSDT fulleride $(U,\mathit\Delta)$ phase diagram we need to evaluate $W$, which is deduced directly from the computed band-structure. The resulting total $t_\mathrm{1u}$ density of states (DOS), roughly shaped in three peaks, and shown in Fig.~\ref{fig-1}(a) resembles that of \ce{NH3K3C60}.\cite{Manini2002} We note that the apparent disagreement between DOS's width and the band-structure splitting at the X point is due to very a small number of states at extreme values and it can thus be seen only in the zoomed view for small DOS values.  We obtain a quite large DOS at the Fermi energy, $N(E_\mathrm{F}) = 17$~states/eV/\ce{C60}, which is a result of expanded lattice structure and the resulting smaller bandwidth $W = 0.5$~eV. For comparison, we refer here to the $W \approx 0.6$~eV reported for \ce{NH3K3C60} and \ce{K3C60}, which are characterized by smaller unit cells.\cite{Manini2002} 

\begin{figure}
\includegraphics[trim = 5mm 7mm 0mm 13mm, clip, width=8.6cm]{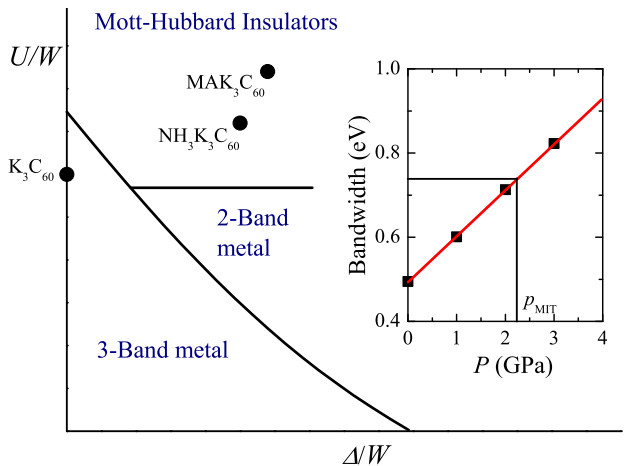}
\caption{MSDT phase diagram for fullerides adapted from Ref.~\citenum{Manini2002} including \ce{MAK3C60} position. Inset: Pressure dependence of the bandwidth computed for the room-temperature \ce{MAK3C60} structure. The horizontal line marks the critical bandwidth 0.74 eV where $(U/W)_c = 1.35$ and where metal-insulator transition is expected.}
\label{fig-2}
\end{figure}

The above results bring forward two interesting aspects of the MA co-intercalation. The first is the position of \ce{MAK3C60} on the MSDT $(U,\mathit\Delta)$ phase diagram. Manini~\textit{et.~al.}\cite{Manini2002} used $\mathit\Delta/W$ as a measure of the anisotropy, which also defines the distance to the MIT boundary in non-cubic fullerides. Based on $\mathit\Delta/W \approx 0.25$ \ce{NH3K3C60} is quite close to both 2- and 3- band metallic phases. It is thus expected that under pressure \ce{NH3K3C60} would almost instantly end up in one of these two metallic states. We find \ce{MAK3C60} to be slightly more anisotropic, judging from  $\mathit\Delta/W = 0.29$, thus deeper in the Mott-insulating phase and further away from the 3-band metallic phase (Fig.~\ref{fig-2}).

Application of a hydrostatic pressure, a standard experimental method for increasing $W$ and thus pushing systems across the metal-insulator boundary, has yet to be tried in this system to verify our calculations. However, based on the above observation, we predict that larger pressures will be needed in \ce{MAK3C60} to access metallic and possibly superconducting states. To estimate the required metal-insulator transition pressure in \ce{MAK3C60} we compute the bandwidth as a function of decreasing unit-cell volume, mimicking the effect of an external pressure. The pressure dependence of the unit-cell parameters was taken from the high-resolution X-ray data measured under hydrostatic conditions.\cite{TakabayashiUP} To prevent unphysical contact between \ce{C60} and MA groups we carried out a structural optimization for each volume. The bandwidth increases monotonically with decreasing unit-cell volume, or increasing pressure, with the slope of $\mathrm{d}W/\mathrm{d}P = 100$~meV/GPa (inset Fig.~\ref{fig-2}). If the effect of pressure could be reduced only to a pressure dependence of $W$, we would predict a steep increase of $T_\mathrm{N}$ in the insulating phase as a function of pressure, estimating for instance $T_\mathrm{N}$(1~GPa)/$T_\mathrm{N}$(0) $\sim$  $(W/W_0)^2 = 1.44$. At higher pressures insulator-metal instability is expected to occur. If we take critical ratio $(U/W)_c = 1.35$ as appropriate for the orthorhombic structure\cite{Manini2002} and a typical value for fullerenes $U = 1$~eV,\cite{Gunnarsson1997} we predict that this transition should take place at around $p_\mathrm{MIT} = 2.3$~GPa. Nonlinear effects on compressibility may push this critical pressure slightly higher. 

Contrasted with the metal-insulator transition, magnetic ordering cannot be understood purely in terms of overall properties such as the bandwidth or crystal-field anisotropy. In particular, the abnormally small $T_\mathrm{N}$ = 11~K of \ce{MAK3C60} as compared with $T_\mathrm{N}$ = 40~K for \ce{NH3K3C60} can hardly be justified by the fairly small differences either in the bandwidth or in the orthorhombic anisotropy. Accordingly, the lattice expansion alone cannot explain the small $T_\mathrm{N}$: the antiferromagnetic ordering must be controlled by other degrees of freedom such as the interaction between co-intercaland molecules and \ce{C60^3-} anions and the related stabilization of a particular JT deformation. We will test this hypothesis in the following sections.

\subsection{The Jahn-Teller effect}
\begin{figure}
\includegraphics[trim = 0mm 0mm 0mm 5mm, clip, width=7.1cm]{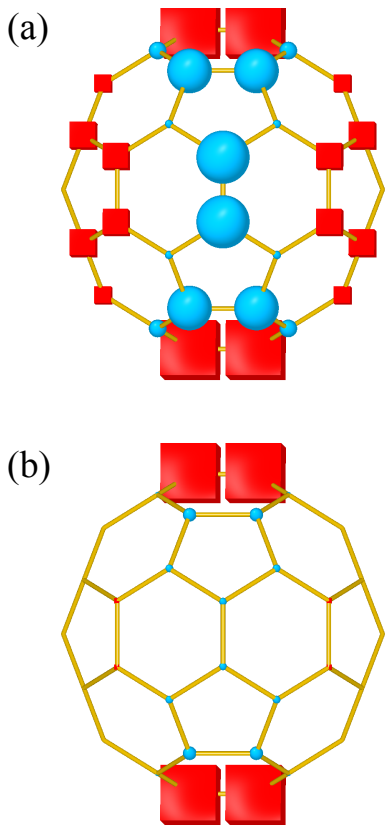}
\caption{Radial molecular distortions of \ce{C60^3-} ions in (a) orthorhombic \ce{C60^3-} expanded structure, (b) \ce{K3C60} orthorhombic environment. The size of blue circles and red squares is proportional to the amount of positive and negative deviations from the mean \ce{C60} radius at a given carbon site, respectively.}
\label{fig-3}
\end{figure}

Let us focus first on JT effect of \ce{C60^3-} anion, which has been predicted\cite{Negri} but has been experimentally much more elusive in fullerene systems.\cite{Khaled} Orthorhombic crystal structures should provide a fertile ground for JT effect investigations. In order to isolate JT effect we consider an artificial enlarged face centered orthorhombic \ce{C60^3-} structure using the room-temperature \ce{MAK3C60} lattice parameters multiplied by a factor of 1.5. To ensure charge neutrality, we add a uniform positive background in the DFT calculation. This positive background is not contributing to the orthorhombic crystal field, therefore any removal of $t_\mathrm{1u}$ degeneracy at the $\Gamma$ point should arise solely from the JT effect on top of the weak residual crystal field from the periodically replicated \ce{C60^3-} ions. The positions of \ce{C60} carbon atoms were relaxed in order to obtain molecular distortions. The \ce{C60^3-} ion deforms spontaneously into a structure with $D_{2\mathrm{h}}$ symmetry [Fig.~\ref{fig-3}(a)]. This is the expected symmetry for $t_{1\mathrm{u}} \otimes H_\mathrm{g}$ JT coupling involving $H_\mathrm{g}$ vibrational modes.\cite{Chancey,Auerbach,Manini1994} Distortions are small with the maximum value of 2~pm. We estimate the energy scale for deformation by realizing that the relaxed structure has $\Delta E_t = 170$~meV lower total energy than the starting structure with undistorted icosahedral \ce{C60}. Former threefold $t_\mathrm{1u}$ degeneracy of the LUMO is now removed, with $t_\mathrm{1u}$ levels split equally by $\sim$50 meV. The lowest $t_\mathrm{1u}$ orbital is doubly occupied, while the third electron goes into the central $t_\mathrm{1u}$ orbital, pinned at the Fermi level. The highest $t_\mathrm{1u}$ orbital is empty. The total energy difference, $\Delta E_t$, has several contributions: the JT effect, the crystal-field effect and bond-length correction due to LDA approximation. To estimate only the JT energy scale we compare the above $\Delta E_t$ with the one obtained for a structural relaxation with equal and fixed occupations of the $t_\mathrm{1u}$ bands, which effectively hinders the JT effect. The difference in the total energy between these two calculations is 57~meV, a typical value for the JT effect.\cite{Chancey,Manini1994} Proper energy scale, typical size of deformations, the right symmetry of \ce{C60^3-} the splitting and occupation of $t_\mathrm{1u}$ orbitals are strong indications that the observed distortion is indeed a result of the JT effect.

The residual crystal field of \ce{C60^3-} ions is immediately seen for smaller lattice parameters, i.e. when room-temperature \ce{MAK3C60} lattice parameters are for instance multiplied by a smaller factor of 1.25. Using the same procedure as above, the structural relaxation ended up with the same deformation of the \ce{C60^3-} molecule. The only difference is that the axis of JT deformation accidentally rotated from the crystal $z$- to $y$-direction. This indicates that we are dealing in both cases with the JT effect and that the change in the JT deformation axis is due to the existence of several equivalent minima of the lowest JT adiabatic potential energy surface (APES).\cite{Chancey} These minima become nearly degenerate for large lattice expansions. On the other hand, when lattice parameters are reduced down toward experimental \ce{MAK3C60} values the orthorhombic crystal field starts to play a role by making some of the APES minima deeper, thus promoting one specific JT deformation.

Adding potassium atoms to the \ce{C60^3-} structure results in an artificial orthorhombic \ce{K3C60} where even stronger orthorhombic crystal field due to the close contact between the K$^+$ and \ce{C60^3-} ions are expected. The structural optimization of the \ce{C60} carbon positions when starting from the JT distorted \ce{C60^3-} atomic positions [Fig.~\ref{fig-3}(a)] leads to deformations of the \ce{C60} molecule shown in Fig.~\ref{fig-3}(b). Molecular deformations are slightly different because the crystal field additionally lifts the $t_\mathrm{1u}$ degeneracy, hence producing pseudo-JT effect. Nevertheless, the resulting symmetry remains $D_\mathrm{2h}$ and the maximal distortions of 2.3~pm are similar to the previous cases. The same holds for the total-energy lowering, $\Delta E_t = 140$~meV. The JT effect is obviously still dominant over orthorhombic crystal field, which represents a smaller contribution to the total energy. The important message of this part is thus that the energy scale of the JT effect is $E_\mathrm{JT} \approx 60$~meV and that for the experimental unit-cell volumes a comparably weak crystal field favors a particular \ce{C60^3-} pseudo-JT deformation.

\subsection{Methylamine - \ce{C60} interaction}

\begin{figure}
\includegraphics[trim = 0mm 0mm 0mm 0mm, clip, width=7.3cm]{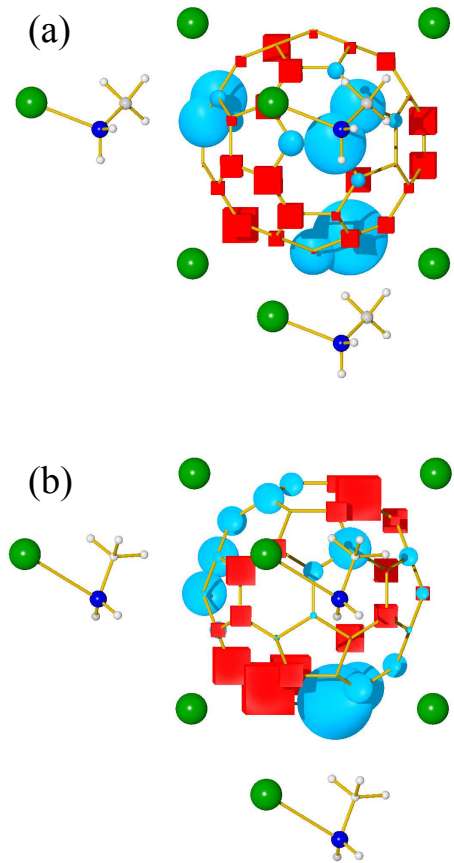}
\caption{\ce{C60} radial deformations of \ce{MAK3C60} for (a) room-temperature structure with relaxed \ce{C60} atom positions only and (b) completely relaxed room-temperature structure. Blue spheres correspond to inward distortions of the \ce{C60} cage and red boxes represent outward cage distortions. The sizes of the markers scale linearly with the amount of distortion.}
\label{fig-4}
\end{figure}

In the \ce{MAK3C60} structure, methyl protons of the MA-K$^+$ groups approach \ce{C60^3-} anions to very short distances and, based on this observation, suggestions about the weak hydrogen bond were formulated in the literature.\cite{Ganin2006} Such close contacts are expected to lead to rather strong crystal fields and, according to the above discussion, to also affect strongly the pseudo-JT effect. It is not \textit{a priori} clear which of the two effects is dominant in this structure. Therefore, \ce{C60^3-} deformations are now investigated in the room-temperature \ce{MAK3C60} structure where only \ce{C60} carbon atoms are allowed to relax. As expected, the presence of MA molecules has a dramatic effect on the \ce{C60^3-} shape [Fig.~\ref{fig-4}(a)]. The maximal cage distortions are significantly larger than in the previous cases - they reach up to 3.4~pm and the total energy is reduced by $\Delta E_t = 303$~meV during the structural optimization. In addition, even the $D_\mathrm{2h}$ symmetry of the distorted \ce{C60} molecule is lost. The maximal distortions are found for carbons facing methyl protons, at a closest-approach distance of 226~pm. Dramatically larger distortions compared to those obtained for \ce{C60^3-} or even for orthorhombic \ce{K3C60} structure indicate that the additional crystal field produced by MA-K$^+$ groups plays a dominant role over the JT effect in \ce{MAK3C60}.

We also address the possible hydrogen-bond formation by relaxing all atomic positions, including those of MA-K$^+$ groups. Hydrogen-bond traces can be detected on the tiny deformation of MA where C-H bond lengths of the \ce{CH3} group are 110.3~pm, 110.4~pm and 110.6~pm, the last one corresponding to the hydrogen with the closest contact to the fullerene molecule. However, with the structural relaxation, the MA-K$^+$ group rotates slightly away from the fullerene molecule, increasing the nearest H-\ce{C60} distance to 228.5~pm, i.e. by 2.7~pm longer than in the experimental structure [Fig.~\ref{fig-4}(b)]. This distance still remains in the typical hydrogen-bond length range, but the \ce{C60^3-} -- methyl proton contact elongation indicates that such bond must be very weak, if it exists at all.

\subsection{Magnetic interactions}

\begin{figure}
\includegraphics[trim = 0mm 90mm 48mm 0mm, clip, width=6cm]{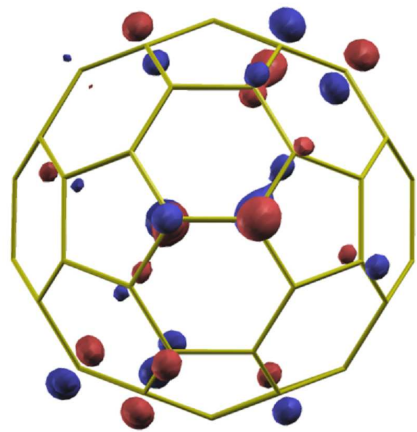}
\caption{A maximally localized Wannier orbital obtained for the \ce{MAK3C60} structure. A isowavefunction surface at levels $\pm$65\% of maximal value are depicted in red and blue color for positive and negative phase, respectively.\cite{Kokalj}}
\label{fig-5}
\end{figure}

The above analyses point toward pseudo-JT effect and strong crystal fields in the \ce{MAK3C60} arising from the MA-\ce{C60^3-} interactions. The remaining question to be addressed is how these effects influence the low-temperature magnetic properties. The appropriate starting point is to construct an effective tight-binding model for the $t_\mathrm{1u}$ bands and then relate the transfer integrals to the exchange coupling constants. To do so, we proceed by switching to the maximally-localized Wannier orbitals (WO).\cite{Mostofi} Setting the energy window to the range around the DFT $t_\mathrm{1u}$ band (Fig.~\ref{fig-1}) we obtain three WOs, each one localized on the same \ce{C60} molecule. Close to the fullerene cage, a typical WO (Fig.~\ref{fig-5}) has the characteristic look of an appropriate combination of carbon $2p_z$ orbitals.\cite{Nomura} The orbital of Fig.~\ref{fig-5} resembles the DFT electron density obtained at the $\Gamma$ point. The distribution of $2p_z$-like orbitals complies with the expected $t_\mathrm{1u}$ symmetry. The other WOs (not shown) have a similar shape, but different orientations. Based on these WOs, we compute their on-site energies and transfer integrals between neighboring \ce{C60} from the Kohn-Sham Hamiltonian as\cite{Mostofi}

\begin{equation}
\label{eq1}
t_{\alpha\beta}(\boldsymbol{R}_i) = \langle w(0,\alpha)|H_\mathrm{KS}|w(\boldsymbol{R}_i,\beta) \rangle.	
\end{equation}

\noindent Here $w(\boldsymbol{R}_i,\alpha)$ is a $\alpha$-th WO and $\boldsymbol{R}_i$ is a lattice translation vector. For $\boldsymbol{R}_i = 0$ Eq.~(\ref{eq1}) gives the on-site WO energies and for $\boldsymbol{R}_i \neq 0$ Eq.~(\ref{eq1}) yields nine transfer integrals for each pair of neighboring \ce{C60} molecules. By diagonalizing the tight-binding model, we verify that the obtained on-site energies and transfer integrals reproduce correctly the DFT electronic band structure as shown by red solid line in Fig.~\ref{fig-1}(a).

We note that the computed on-site tight-binding Hamiltonian has non-zero off-diagonal elements, indicating hybridization within the obtained set of WOs. By changing basis to an equivalent one where the on-site Hamiltonian becomes diagonal, we derive a new set of WOs with well-defined on-site energies. There are two indications that the new set of WOs is appropriate for our purposes: (i) the new WOs still reproduce band structure and (ii) in the Mott-insulating limit such WOs are a good approximation for the localized $t_\mathrm{1u}$ electronic orbitals with well-defined on-site energies, in our case separated by 32 and 67~meV.

To estimate the exchange coupling constants we focus only on the transfer integrals of the second WO, the half-filled one at intermediate energy. The neglect of the other (inter-band) hopping elements between neighboring \ce{C60} sites amounts to neglecting inter-band transitions. This selection can be justified by noting that (i) the retained WO is mainly responsible for the DOS at the Fermi level [see Fig.~\ref{fig-1}(a)] and (ii) in a picture where the Coulomb repulsion is the dominating interaction, the intermediate orbital hosts the unpaired spin in the spin-$1/2$ state of each \ce{C60^3-} molecular unit, with lower fully occupied and upper empty orbital. Within this approximation we estimate the interfullerene exchange coupling constants with Hubbard's expression $J_{ij} = 4t_{ij}^2/U$, where we take $U = 1$~eV. For the considered \ce{MAK3C60} structure the strongest exchange couplings $J_3 = J_4 = 1.9$~meV are found for nearest \ce{C60} neighbors along the (0 1 1) and (1 0 -1) directions. We stress that these $J$'s are of the right order of magnitude as estimated from the measured Curie-Weiss temperature $\mathit\Theta = 86$~K.\cite{Ganin2007}  All other exchange constants are much weaker, not exceeding $0.12 \cdot J_3$, see Table~\ref{table-1}.

\begin{table}
\caption{Exchange coupling constants for the experimental room-temperature structure [Fig.~\ref{fig-4}(a)] and the relaxed structure [Fig.~\ref{fig-4}(b)]. See text for details.}

\label{table-1}
	\begin{tabular}{|c|c|c|}
	\hline
	\textbf{Neighbors} & $\mathbf{J}_\textbf{Exp. structure}$ \textbf{(meV)} & $\mathbf{J}_\textbf{Relaxed}$ \textbf{(meV)} \\
	\hline
	1: (1 1 0) & 0.03 & 0.06 \\
	\hline
	2: (1 0 1) & 0.02 & 0.02 \\
	\hline
	3: (0 1 1) & \textbf{1.94} & 0.01 \\
	\hline
	4: (1 0 -1) & \textbf{1.89} & \textbf{3.07} \\
	\hline
	5: (0 1 -1) & 0.00 & 0.73 \\
	\hline
	6: (1 -1 0) & 0.22 & 0.72 \\
	\hline
	\end{tabular}
\end{table}

The obtained exchange network indicates a quasi-two-dimensional magnetic structure of \ce{MAK3C60}. Low dimensionality could account for the reduced experimental N\'eel temperature. However, although this is certainly a viable possibility, we stress that this conclusion is based on the high-temperature crystal structure. Since we proved that the interaction between MA and \ce{C60^3-} is the governing factor for pseudo-JT effect, we stress at this point that the low-temperature exchange network may in fact be different. How sensitive interfullerene exchange interactions are on the precise position of MA-K$^+$ groups becomes immediately evident when we consider relaxed DFT structure with a slightly rotated MA-K$^+$ groups [Fig.~\ref{fig-4}(b)]. For this structure the strongest exchange is $J_4 = 3$~meV, but now along (1 0 -1). The quasi low-dimensional picture seems to still hold since the next two strongest interactions are much weaker, i.e. $J_5 = 0.23 \cdot J_4$ (0 1 -1), and $J_6 = 0.23 \cdot J_4$ (1 -1 0). All other exchange interactions are negligible (Table~\ref{table-1}). We stress that in all these structures the considered MA-K$^+$ order is ferro-orientational. Any deviation from this MA-K$^+$ configuration, for example antiferro-orientational MA-K$^+$ order as the candidate for the low-temperature \ce{MAK3C60} structure, could significantly change the quasi low-dimensional character of our system. Moreover, if a certain degree of disorder in MA-K$^+$ orientations is present, then one would anticipate also a distribution of exchange coupling constants, and an even smaller ordered moment. That would explain the static magnetic order with a broad-local-field distribution measured by $\mu$SR.\cite{Takabayashi2006} This situation is somehow reminiscent of ferromagnetic TDAE-\ce{C60}, where disorder in JT deformations has been argued to be responsible for the reduced order parameter\cite{Mihailovic,Blinc,Jeglic2003} and broad distribution of local magnetic fields also measured in $\mu$SR experiment, for instance.\cite{Lappas1995} We conclude that pseudo-JT orientation is an essential factor in interfullerene exchange coupling in \ce{MAK3C60}. Judging from the experimental high-temperature structure, \ce{MAK3C60} may even show low-dimensionality effects thus explaining the reduced $T_\mathrm{N}$. However, other pseudo-JT orientations may be active at low temperatures and even suppression of $T_\mathrm{N}$ due to the disorder effects cannot be excluded at this stage. Further progress in understanding of magnetic properties of \ce{MAK3C60} would be possible only when the low-temperature crystalline structure of \ce{MAK3C60} is known.

\section{Conclusions}
Antiferromagnetic Mott-Hubbard insulating \ce{MAK3C60} had been studied by means of DFT calculations. We found the pronounced orthorhombic crystal field anisotropy in the electronic structure and judging from $\mathit\Delta/W = 0.29$ \ce{MAK3C60} should be deep in the insulating part of the phase diagram. Comparisons between different orthorhombic structures demonstrate that the JT effect is present and that its energy is around 60~meV. However, the interaction between cointercalated MA molecules and \ce{C60^3-} anions is comparatively strong in \ce{MAK3C60}, leading to a strong crystal field and pseudo-JT effect stabilized at $\sim$300~meV. The presence of strong pseudo-JT effect should be responsible for the suppression of $T_\mathrm{N}$. Calculations based on the high-temperature structure with imposed ferro-orientational order of MA-K$^+$ groups suggest low-dimensionality in magnetic exchange; however other explanations such as a disorder in pseudo-JT cannot be excluded at this stage.
A similar interplay of Mott and JT physics is believed to play an important role in the superconductivity of underdoped cuprates. In the present study we emphasize the similar importance of JT effect in strongly correlated \ce{C60^3-} structures and setting it as one of the most important parameters for the high-temperature superconductivity in fulleride salts.

\acknowledgments 
We thank Gianluca Giovannetti for many stimulating discussions about these DFT results. This work was supported in part by the Slovenian research agency through project No. J1-2284. E.T. and D.A. acknowledge the financial support by the European Union FP7-NMP-2011-EU-Japan project LEMSUPER under contract no. 283214. 

\appendix

\end{document}